\pdfoutput=1
\documentclass[12pt,twoside,draft]{IEEEtran_v15} 
\usepackage{graphicx,cite,amsmath,amssymb,hhline,pbox,array,longtable}
\begin{document}

\title{A Survey of Channel Modeling for UAV Communications }

\author{
\smallskip {\normalsize $\mbox{Aziz Altaf Khuwaja}^{}$, $\mbox{Yunfei Chen}^{}$, {\em Senior Member, IEEE,} $\mbox{Nan Zhao}^{}$, {\em Senior Member, IEEE,}}\\ \newline
\smallskip{\normalsize $\mbox{Mohamed-Slim Alouini}^{}$, {\em Fellow, IEEE,} $\mbox{Paul Dobbins}^{}$ }
\thanks{Aziz Altaf Khuwaja
    is with the School of Engineering,
    University of Warwick, Coventry, U.K. CV4 7AL and also with the Department of Electrical Engineering,
    Sukkur IBA University, Sukkur, Sindh, Pakistan  
    (e-mail: A.khuwaja@warwick.ac.uk)}\\
\thanks{Yunfei Chen
    is with the School of Engineering,
    University of Warwick, Coventry, U.K. CV4 7AL
    (e-mail: Yunfei.Chen@warwick.ac.uk)}\\ 
\thanks{Nan Zhao 
    is with the School of Information and Communication Engineering,
    Dalian University of Technology, China
    (e-mail: zhaonan@dlut.edu.cn)}\\
\thanks{Mohamed-Slim Alouini
    is with the King Abdullah University of Science and Technology (KAUST),
    Thuwal 23955-6900, Makkah Province, Kingdom of Saudi Arabia
    (e-mail: slim.alouini@kaust.edu.sa)}\\ 
    \thanks{Paul Dobbins
    is with the Telent,
    Haywood Road, Warwick, CV34 5AH, UK
    (e-mail: paul.dobbins@telent.com)}\\   
    }
\thispagestyle{empty} \setcounter{page}{0}
\maketitle{
\renewcommand{\baselinestretch}{1.73} \normalsize
}
\begin{abstract}
Unmanned aerial vehicles (UAVs) have gained great interest for rapid deployment in both civil and military applications. UAV communication has its own distinctive channel characteristics compared with widely used cellular and satellite systems. Thus, accurate channel characterization is crucial for the performance optimization and design of efficient UAV communication systems. However, several challenges exist in UAV channel modeling. For example, propagation characteristics of UAV channels are still less explored for spatial and temporal variations in non\textendash stationary channels. Also, airframe shadowing has not yet been investigated for small size rotary UAVs. This paper provides an extensive survey on the measurement campaigns launched for UAV channel modeling using low altitude platforms and discusses various channel characterization efforts. We also review the contemporary perspective of UAV channel modeling approaches and outline some future research challenges in this domain.
\end{abstract}

\begin{keywords}
Channel characterization, channel models, measurement campaigns, UAV communication.  
\end{keywords}

 
\section{Introduction}
\label{one}
Unmanned aerial vehicles (UAV) aided communication has seen drastic development in a variety of applications. For instance, it can be used in military operations for border surveillance and information gathering in hostile environment. Also, it can be deployed to monitor civil emergency services, conduct humanitarian missions and facilitate scientific data collection.  Most of these applications deploy UAVs as low altitude platforms, and they can be fully autonomous or remotely operated. In order to ensure safety and high reliability, it is of utmost importance to thoroughly characterize the UAV channels particularly for low altitude platforms. Many research organizations and standardization bodies have worked together to establish pragmatic UAV frameworks. For example, special committee (SC\textendash228) has been formed by the Radio Technical Commission for Aeronautics (RTCA) in 2013 to frame minimum performance standards for UAV operations \cite{rcta1}. RTCA has also established drone advisory committee in 2016 to ensure safe introduction of UAVs into the national airspace system \cite{rcta2}. Also, National Aeronautics and Space Administration (NASA) and Federal Aviation Administration (FAA) have launched a joint research initiative to integrate UAVs in national airspace system across the United States \cite{nasa}.\\
The most unique features that distinguish UAV communication systems from conventional wireless communication systems or the unique channel characteristics of UAV communications are:
\begin{itemize}
	\item the distinct communication channels, i.e., air\textendash ground (AG) and air\textendash air (AA) channels
	\item the spatial and temporal variations in non\textendash stationary channels
	\item the airframe shadowing caused by the structural design and rotating capability of UAV. 
\end{itemize}

In a diverse propagation environment where the UAV operates, these features become more challenging. Propagation characteristics for terrestrial cellular systems are often corroborated with well\textendash established empirical and analytical models. The satellite links for land mobile systems have also been thoroughly investigated in the literature \cite{Panagopoulos},\cite{Chini}. However, these models are often not well suited to characterize the UAV propagation channel due to the aforementioned unique attributes of theirs.
   
On the other hand, reliable analytical models are necessary to evaluate the performance of different wireless techniques and to provide assistance to link budget calculations. In the context of AG channels in UAV communications, modeling approaches can be generally classified in three categories. The first approach is to develop deterministic models using environmental parameters, while considering the UAV altitude and elevation angle from the ground. Such models are useful to study the fading effects in the channel \cite{Daniel}, \cite{Feng01}, the propagation conditions \cite{Feng02},\cite{hourani01} and hence can provide coverage analysis for optimal UAV position \cite{hourani02}, \cite{hourani03}. The second approach is to develop the tapped delay line (TDL) model to characterize the direct path as well as the multipath components. This gives the wideband frequency\textendash selective parameters derived from channel impulse response \cite{Matolak03}\textendash\cite{Matolak05}. This approach is particularly important if non\textendash stationarity in AG channel is to be addressed. Finally, the geometric\textendash based stochastic model is desirable for evaluating spatial\textendash temporal characteristics in a geometric simulation environment. This approach is more preferable to characterize the AG channel in 3D plane with less environmental parameters \cite{newhall01}\textendash\cite{zeng}.

However, analytical models alone do not always describe the real behavior of the propagation channel, because of deficient realistic assumptions. Therefore, empirical studies initiated by measurement campaigns are essential. Most of the works reported in the literature \cite{holzbock}\textendash\cite{Mrice04} are pertinent to the AG channel characterization based on the measurement campaigns launched with manned aircrafts in high altitude platform. However, these findings cannot be directly applied to the single\textendash hop UAV network deployed in low altitude platform, i.e., up to 120 m as permitted by FAA in USA \cite{faa} and Civil Aviation Safety Authority (CASA) in Australia\cite{casa}. It is evident from the studies in \cite{Simunek01}\textendash\cite{hourani} that the impact of airborne platforms are significant for the channel characteristics of UAV communications. Moreover, less research efforts have been made to tackle with the shadowing induced in AG and AA channels by the UAV structural design and maneuvering. In addition, the wide sense stationary uncorrelated scattering (WSSUS) assumption may be violated in some UAV\textendash aided applications. Thus, in order to avoid exaggerated performance evaluation from the analytical channel model, it is important to estimate the fading statistics with stationary intervals. The AA channel in multi\textendash hop UAV network has been empirically characterized in the literature with low power radios based on IEEE 802.15.4 \cite{allred}\textendash\cite{ahmed01} and IEEE 802.11 standards \cite{goddemeier02}\textendash\cite{yanmaz03}. But these studies only reported large\textendash scale fading statistics and the impact of antenna orientation on channel performance, thus the Doppler spectrum for AA channel is understudied.

Despite the importance of channel modeling in UAV communications, very few survey studies are available in the literature. For instance, reference \cite{gupta} identified key issues related to the formation of multi\textendash UAV network, but this survey focuses more on the communications and especially the control of UAV. Also, aerial networking characteristics and requirements are reviewed in \cite{shayat} for civil applications. However, this survey mainly discussed the communications aspects of UAV, in particularly network layer designs. Both \cite{gupta} and \cite{shayat} have very light touch on the channel modeling. On the other hand, the physical layer characterization of the AG channel at L and C bands was comprehensively reviewed in \cite{Matolak02}. However, measurement campaigns in this paper were reported mainly for aeronautical communications and land mobile satellite systems at the L and C bands. In contrast, our survey will review the current advances for the UAV channel characterization.

The rest of the paper is organized as follows. In Section II, we will review the measurement campaigns launched with UAVs as low altitude platforms, where we will categorize them according to the bandwidth of channel sounder, the low\textendash power and low\textendash cost radios and the widely deployed cellular infrastructure. Characterization of AA and AG propagation using empirical channel models will be discussed in Section III. In Section IV, we will categorize analytical UAV channel models as deterministic, stochastic and geometry\textendash based stochastic. In Section V, we will highlight some important issues pertinent to the airframe shadowing, non\textendash stationary channels and the applicability of diversity techniques in UAV communications. Finally, we will discuss future research challenges in UAV channel modeling.

\section{Measurement Campaigns}
\label{two}

Propagation channel models developed using analytical approaches do not always give satisfactory performances in real\textendash time deployment due to inadequate or unrealistic assumptions. In this case, the actual behavior of the propagation channel can only be understood via field measurements. A number of measurement campaigns have been carried out in diverse environments to understand the UAV channel characterization. Some of these campaigns analyze the AG propagation channel with large\textendash scale and small\textendash scale fading effects and hence only propose empirical channel models. Other measurement campaigns assess the use of diversity techniques for range extension and enhancement of communication channel throughputs.

In the literature, most of the measurement campaigns have been conducted using two types of aerial vehicles. The first type of aerial vehicles are small and medium sized manned aircrafts. For instance, in \cite{Matolak03}\textendash\cite{Matolak05}, S\textendash3B Viking aircraft was used to comprehend the AG channel characteristics at L and C bands in different environments. In \cite{Jchen}, the Cessna-172S aircraft was used to evaluate the performance of a $4\times4$ MIMO enabled OFDM system for the AG channel. In \cite{Mrice01} and \cite{Mrice02}, UH\textendash1H military helicopter was used to study the AG channel in a $4\times2$ MIMO configuration to achieve the diversity gain and equalization efficacy to mitigate ISI in frequency\textendash selective channels. In \cite{Yjiang}, a news\textendash reporting helicopter was used to attain spatial multiplexing gain and throughputs for airborne communication in $2\times2$ MIMO settings. The logistics involved in the measurement campaigns using manned aircrafts are expensive and daunting. Therefore, the second type of aerial vehicles i.e., UAVs are preferable to reduce the cost. In this case, the UAV payload is often integrated with an on\textendash board processor to control flight dynamics and wireless equipment to collect measurement data. In addition, the experimental setup also contains antennas to radiate and receive RF signals, GPS system to record telemetry data and inertial measurement unit (IMU) to measure flight dynamics such as pitch, yaw and roll angles. In the rest, we will mainly focus on the measurement campaigns using UAVs and will categorize them into three main groups based on the wireless techniques. The first group of the measurement campaigns use narrowband or wideband channel sounder. The second set of measurement campaigns use IEEE 802.11 radios. In the third group, the measurement campaigns are accomplished using the widely deployed cellular network infrastructure.

\subsection{Narrowband and Wideband Channel Sounder}
\label{two-one}
\subsubsection{Narrowband Measurement Systems}
\label{two-one-one}
They evaluate the Doppler frequency shift and the channel gain experienced by a narrowband continuous wave (CW) signals. These systems have a channel sounder that generates pilot tones at a single carrier frequency using a CW generator. Examples of narrowband measurement campaigns for characterizing the AG propagation channels in aeronautical communications for very high frequency (VHF) band are \cite{Wvergara},\cite{Chamberlin}, for L band \cite{Child} and for higher frequency (HF) band \cite{Mrice03}.

In \cite{Simunek01}, the measurement campaign was performed in an urban area of Prague, Czech Republic, using 2 GHz CW transmitter with a bandwidth of 12.5 kHz. The measurement setup includes an airship UAV mounted with transmitter and four\textendash channel custom\textendash made receiver located at the ground station. Also, monopole antennas were used at both the transmitter and receiver sides. The UAV flew between 100 to 170 m above the ground level at low elevation angle between 1$^{\circ}$ to 6$^{\circ}$. The authors have statistically characterized the AG channel which fits in between a purely terrestrial link and a land mobile satellite system. They have also presented a narrowband channel estimator capable of replicating signal dynamics. Some related measurement campaigns were conducted with similar equipment in Prague for path loss model in urban area \cite{Simunek02} with a flight altitude between 150 to 300 m. Further, measurements in \cite{Simunek03} and \cite{Simunek04} were conducted in urban and wooded areas, respectively, to study space diversity techniques for similar situations.

In \cite{Cai}, field experiments were performed in a suburban terrain of Madrid, Spain, with two frequency bands i.e., 5.76 GHz for narrowband measurements. Field trials were carried out using a hexacopter UAV, universal software defined radio peripheral (USRP) hardware, and cloverleaf antennas with circular polarization. In the narrowband measurement scheme, frequency modulated CW transmitter was used for both vertical and horizontal flight routes in order to characterize the large\textendash scale variation. In this case, USRP was used as the ground receiver. The UAV flew at an altitude between 0 to 50 m for the vertical flight route and covered distance of 20 m and 30 m for horizontal route. The authors have investigated large\textendash scale fading effects in UAV propagation channel and computed path loss exponent for both vertical and horizontal flight direction using the dual slope and the log\textendash distance path loss models, respectively. They found that for the vertical flight direction, the attenuation decreased below the breakpoint distance and then increased with UAV altitude. Whereas, the attenuation increased exponentially with the horizontal flight direction. Furthermore, they have also modeled fast fading effects with Rician distribution.

Narrowband measurement systems are appropriate for computing frequency non\textendash selective fading parameters because of the limited channel sounding spectrum. However, these systems may not be preferred in rich multipath environment to characterize the performance of UAV propagation for coherence bandwidth and multipath delay statistics. Also, the performance evaluation of MIMO capacity using these systems may seems to be a difficult task as they are not able to resolve individual multipath components.
           
\subsubsection{Wideband Measurement Systems}
\label{two-one-two}
They determine the channel impulse response (or transfer function) and frequency\textendash selective parameters derived from it. Wideband channel measurements for characterizing the aeronautical propagation channels are mostly conducted with spread spectrum channel sounder. One such type is correlative channel sounder, where pseudonoise (PN) sequence is transmitted as the channel sounding signal and the received signal is then correlated at the receiver with the same PN sequence. For example, correlative channel sounder was used in \cite{newhall} and \cite{Mrice04} for measuring multipath effects. In the context of characterizing the UAV propagation channel, wideband frequency\textendash selective parameters are often measured with USRP platforms, for instance, as used in \cite{Cai} and \cite{gutierrez}. This platform provides more flexibility in terms of low\textendash power consumption and multiple frequency bands.
 
In \cite{Cai}, the wideband measurement campaign was also performed with the channel sounding signal generated by the LTE base station at the frequency of 1.817 GHz. In this case, USRP was mounted on the UAV as the receiver module with antennas placed below the UAV propellers. In this work, the small\textendash scale variations in the UAV propagation channel was characterized with the measured channel impulse responses and estimated delay spread and power delay profile. The authors have analyzed channel statistics using the cumulative distribution function (CDF) and observed the random behavior of the multipath components at different UAV altitude.
 
In \cite{khawaja}, the measurement campaign was conducted for open and suburban spaces on the campus of Florida International University. The experimental setup consisted of a quadcopter UAV, the ultra\textendash wideband (UWB) channel sounding radio tuned at the frequency of 4.3 GHz with a range between 3.1 GHz to 5.3 GHz and the planar elliptical dipole antennas. In this setting, UAV altitude raised from 4 m to 16 m with the step size of 4 m. The same UWB radio was used as the elevated ground receiver at two different heights and positioned in three different scenarios. In the first scenario, the receiver was placed under the tree canopy and elevated at 1.5 m from the ground. In the second scenario, the receiver was placed at the same height with clear line\textendash of\textendash sight (LOS) to the transmitter. In the third scenario, the receiver was laid down at 7 cm from the ground in LOS condition. In this work, the authors have characterized AG propagation channel, where they proposed the empirical path loss model for both static and mobile UAVs. They found the worst path loss attenuation for mobile UAV motion in the first scenario, whereas, the best for static UAV in the second scenario. Also, they characterized the fading channel as Nakagami m and presented multipath propagation model.
 
In \cite{gutierrez}, the measurement campaign was performed in a residential area and mountainous desert landscape in Arizona, USA, with SDR platform tuned at 5.8 GHz. The USRP radio was attached with a octocopter UAV and also served as ground base station, where both radios were equipped with the dual band vertically oriented omnidirectional antenna and controlled by open source GNU radio. The authors have characterized the frequency\textendash selectivity of the AG propagation by the average and RMS delay spread of the channel. Also, Doppler power spectrum was calculated by summing the entire range of scattering function delay. They analyzed channel statistics with the CDF and found that the desert terrain causes substantial delay spread in the AG propagation than in a residential area. Moreover, CDF analysis followed a log\textendash normal trend for the RMS Doppler spread.
 
Wideband measurement campaigns are desirable to evaluate both narrowband and wideband frequency\textendash selective channel parameters. However, additional computational capabilities are required to process the raw data collected from the measurements. Therefore, this type of measurement systems may not be suitable for real\textendash time characterizing of fading channel parameters. Also, cost and physical dimensions of wideband channel sounding equipment are other possible constraints that need to be considered. 
 
 \subsection{IEEE 802.11 based UAV Measurements}
\label{two-two}
UAV channel characterization using commercial off\textendash the\textendash shelf 802.11 radios are desirable due to their low power consumption, cost effectiveness and flexibility to be integrated with small size UAVs. However, the performances of such radios are prone to interference and background noise. Also, fixed narrowband frequency and limited communication range are other constraints to evaluate fading channel parameters. Channel characterization efforts reported in the literature for multi\textendash hop UAV networks were based on IEEE 802.11 in \cite{frew}\textendash\cite{hayat} and also IEEE 802.15.4 ZigBee devices in \cite{allred}\textendash\cite{ahmed01}. In this section we review the measurement campaigns relevant to 802.11 radios for single\textendash hop UAV network only.

In \cite{goddemeier02}, the measurement campaign was performed in laboratory and outdoor environment to particularly study the altitude\textendash dependent multipath propagation in AA channel. The measurements were collected with 802.11 a/b/g/n WLAN devices from two different vendors and deployed in three outdoor scenarios using a hexacopter UAV. The laboratory experiments were conducted for sensitivity analysis and calibration purpose. In the first scenario, the impact of flight distance followed a free space path loss model. In the second scenario, good signal reception was attained between 170$^{\circ}$\textendash230$^{\circ}$ and the worst signal for a yaw angle of 240$^{\circ}$\textendash260$^{\circ}$. Finally the effect of the ground reflected multipath components on UAV altitude was examined for the flight altitude between 10 to 40 m and proposed the height dependent Rician model with K factor reliant on the UAV altitude.

In \cite{yanmaz03}, the measurement campaign was performed in an open space using a quadcopter UAV and an access point (AP) connected with 802.11a WLAN interface. Also, IMU module was used to measure the UAV position and orientation in 3D plane. Measurements were collected with three horizontally aligned dipole antennas and at the flight altitude between 15 to 110 m. The authors observed that for both AG and AA channels, the path loss exponent computed by log\textendash distance model roughly matched with that of free space propagation. Also Nakagami m distribution was found to be a good fit for a multipath fading channel. Furthermore, inter\textendash arrival time of packet and retransmission attempts were analyzed by empirical CDFs. 

Related field trials were conducted in \cite{yanmaz02} for an open space and a campus environment for the UAV flight altitude varies between 20 to 120 m in different testing scenarios. Two vertically polarized omnidirectional antennas were mounted on both UAV and AP. The authors found that the optimal antenna orientation can alleviate the impact of UAV altitude on received signal strength and throughputs. Moreover, horizontally aligned antennas reduced the affect of UAV yaw angle on throughputs. They have also found that the propagation condition followed that in the free space for an open field.

In \cite{cheng}, the measurement campaign was launched at a private airfield in Connecticut, USA, using a 802.11a radio mounted on a fixed\textendash wing UAV. Commercially available dual\textendash band omnidirectional antenna and custom\textendash made antennas tuned at 5.28 GHz were tested with 32 orientation pair configurations. The UAV flew approximately at 64 km/h airspeed and maintained an altitude of 46 m over the ground receiver nodes. The authors evaluated the throughput reliance with UAV transmit antenna and reported highest rates with horizontal dipole, orthogonal to flight direction and parallel to the ground. In addition, they also estimated that path loss roughly followed free space propagation.

In \cite{hague}, a related measurement campaign was performed with both 2.4 GHz 802.11g and 5.8 GHz 802.11a devices. In this case, the authors computed the maximum range attained with 802.11a radio and compared that by 802.11g. They found that 802.11g node can provide robust communication at the altitude of approximately 183 m.  In this work, another experimental trial was conducted with 900 MHz 802.11 radio to determine received signal strength and throughput performances. They found significant communication range up to 2000 m with throughputs in Mbps. In addition, they have analyzed the path loss attenuation with linear regression method. In \cite{Hkung}, the measurement campaign was done in the farmland area amid by woods. In this work, AG channel characterization was performed in terms of network level diversity gain and found significant enhancement in packet transmission rate by multiple receivers.

Low\textendash power and cost\textendash effective IEEE 802.11 radios are preferable for narrowband field measurements in UAV networking. Also this platform provides an opportunity for characterizing the UAV propagation channel with various antenna orientations. Thus, it provides the optimal placement and alignment of on\textendash board UAV antennas. However, in a complex communications environment where UAV operates, radio interference from other 802.11 equipment can be challenging. In this case, one possible solution is to maintain high signal\textendash to\textendash noise\textendash plus\textendash interference ratio (SINR) at the physical layer for each aerial link in the presence of possible interference from the adjacent radio device.

\subsection{Cellular\textendash Connected UAV Measurements}
\label{two-three}
Cellular networks can be considered as a prospective candidate to facilitate UAV applications in civil and commercial domains. Widely deployed cellular infrastructure can be utilized to provide reliable AG channels and hence, cut the cost of investing additional ground infrastructure and spectrum allocation. However, since cellular\textendash connected UAVs depend on the cellular network and cellular infrastructure can collapse due to natural disaster, a viable fail\textendash safe mechanism is needed. Other challenges, such as down\textendash tilted base station antennas, neighboring cell interference, handover performance, multiple access, UAV mobility and link security, also need to be addressed thoroughly before the widespread implementation of UAV network connected to the cellular networks. This has motivated several mobile operators, telecommunication vendors and research organizations to further scrutinize the propagation channel characteristics between cellular base station and airborne UAV. For example, Qualcomm Technologies has launched field measurements in San Diego, California, to assess the LTE network performance in low altitude platform using quadcopter UAV \cite{qualcomm}. In another example \cite{ericsson}, Ericsson and China Mobile have conducted measurement trials in China's Jiangsu province to develop 5G prototype enabled by drone UAVs.

In \cite{goddemeier03}, the measurement campaign was launched in urban and rural scenarios in Germany to characterize the propagation channel between UAV and cellular base station, using 900 MHz GSM network and 1.9\textendash2.2 GHz UMTS services. Field measurements were carried out with the fixed\textendash wing UAV and captive balloon at the altitude up to 500 m. This work evaluated the overall aerial performance in terms of received signal strength and handover analysis in both urban and rural scenario. The authors have made assumptions that the attenuation was independent from frequency and distance. It was found that due to signal degradation at higher UAV altitude the availability of base station decreases. To conclude, a good RF coverage was achieved in a rural environment due to less ground obstacles than in urban terrain.

In \cite{tavares} and \cite{afonso}, measurement campaigns were launched under the SAAS project (remote piloted semi\textendash autonomous aerial surveillance system using terrestrial wireless networks) in an urban environment of Lisbon, Portugal to investigate the applicability of terrestrial cellular networks in UAV communication. In \cite{tavares}, the field trials were performed at GSM, UMTS and LTE cellular bands using spectrum analyzer and antenna tied with meteorological balloon, deployed as UAV platform. Received signal power was recorded from the roof mounted base station at the UAV altitude between 11 to 18 m. In this work, empirical model was obtained for path loss attenuation in outdoor urban scenario. The worst case scenario was reported due to the radiation pattern of the base station antenna where received signal power dropped as UAV climbed above the base station height. However, handover analysis was not studied in this scenario. On the other hand, reference \cite{afonso} presented multi\textendash UAV network architecture based on cellular and IP networks. They have assessed the network level performance with quality of service measurements in terms of received power, latency and jitters.
     
In \cite{amorim}, the measurements campaign was performed in the rural environment at 800 MHz LTE networks with two different cellular service providers in Denmark. Two flight zones 7 km apart from each other were demarcated for the experimental site, surrounded by multiple base stations with a height between 20 to 50 m. The UAV flew in circular track of 500 m diameter at the airspeed of 15 km/h and varies altitude between 15 to 120 m. The authors have found considerable reduction in path loss exponent and shadowing variation as UAV altitude increased. Therefore, their findings exhibited that the UAV propagation channel necessitates altitude dependent parameters for channel modeling.   

In \cite{hourani}, the measurement campaign was launched for suburban terrain in Victoria, Australia, at 850 MHz LTE cellular network, using a quadcopter as UAV and Andriod based mobile phone was used for logging samples of received signal. The experimental site covered 12 $km^{2}$ surrounded area with a single base station of height 30 m. UAV flew at an average speed of 17 km/h at the altitude between 15 to 120 m. From the measurement result, they obtained angle\textendash dependent parameters to characterize propagation channel between cellular base station and UAV in airborne platform.

In \cite{teng}, another measurement campaign was conducted for an open area and mock village in California, USA, at 909 MHz cellular band. Measurement setup consisted of a quadcopter UAV equipped with sensor package, transmitting radio with 13.9 m pneumatic mast serving as base station and ground controller. The UAV flew at the altitude between 40 to 60 m and followed linear and radial flight patterns. In this work, the authors have proposed the compositional path loss model to account two\textendash ray ground reflection propagation and diffraction losses. Also, they identified low coverage zones in cellular\textendash connected UAV networks for beyond LOS operations and named this phenomena as \textquotedblleft holes in the sky\textquotedblright. They pointed out that the primary causes for this phenomena are interference caused by two\textendash ray ground reflection, diffraction losses incurred by the fresnel zone of propagation path and nulls in the antenna radiation pattern.

Cellular networks seems to fulfill future requirements of UAV communication as they provide extended coverage to the large area via handover between multiple base stations. But cellular networks are not designed to provide AG propagation above the base station height due to down\textendash tilted sector antennas. Also, UAV applications such as search and rescue services and disaster management may be suffered due to infrastructure failure. In this case, aerial heterogeneous network can be a promising fail\textendash safe framework for enabling coexistence between terrestrial communication networks and satellite systems to ensure redundancy of UAV communications. In this section, we have reviewed the measurement campaigns using UAVs as low altitude platforms. In Table I, we summarize the aforementioned measurement campaigns.

\begin{table}[tp] \footnotesize
\renewcommand{\arraystretch}{1.3}
\caption{Measurement Campaigns}
\label{Measurement Campaign}
\begin{tabular}{|p{1cm}|p{2cm}|p{1.8cm}|p{1.9cm}|p{1.8cm}|p{4.5cm}|}
  \hline
  \textbf{Ref.} & \textbf{Frequency} & \textbf{UAV} & \textbf{Scenario} & \textbf{Altitude} & \textbf{Channel Statistics} \\
  \hline
  \cite{Simunek01} & 2 GHz & Airship & Urban & 100-170 m & PDF, CDF, AFD, LCR, PSD, AF\\
  \hline
  \cite{Simunek02} & 2 GHz & Airship & Urban & 150-300 m & PL \\
  \hline
  \cite{Cai} & 5.76 GHz & Hexacopter & Suburban & 0-50 m & PL, SF, K, RMS, CDF \\
& 1.817 GHz & & & & \\  
    \hline
    \cite{khawaja} & 4.3 GHz & Quadcopter & Open field, & 4-16 m & PL, SF, $\mu$, $\xi$, PDF, CDF, RMS\\
     & & & Suburban & &BC \\
     \hline 
 \cite{goddemeier02} & 2.4 GHz & Hexacopter & Laboratory, outdoor & 10-40 m & PL, PAS, K, PDF \\
  \hline
  \cite{yanmaz03} & 802.11a & Quadcopter & Open field & 15-110 m & PL, PAS, CDF \\
    \hline
 \cite{yanmaz02} & 802.11a & Quadcopter & Open field, campus area & 20-100 m & PL \\
 \hline
  \cite{cheng} & 802.11a & Fixed\textendash wing & Airfield & 46 m & PL \\
  \hline
  \cite{hague} & 802.11a/g, & Fixed\textendash wing & Airfield, Rural & 46 m, & PL \\
  & 900 MHz &  & &107-274 m & \\
  \hline
  \cite{goddemeier03} & GSM, UMTS & Fixed\textendash wing, captive balloon & Urban, rural & 0-500 m & PL \\
   \hline
 \cite{tavares} & GSM, UMTS, LTE & Weather balloon & Urban & 11-18 m & PL\\
 \hline
 \cite{amorim} & LTE (800 MHz) & Hexacopter & Rural & 15-100 m &  PL, SF\\
 \hline 
  \cite{hourani} & LTE (850 MHz) & Quadcopter & Suburban & 15-120 m &  PL, SF\\
  \hline
   \cite{Simunek03},\cite{Simunek04} & 2 GHz & Airship & Urban, wooded & 100-170 m & CDF, DG, AFD, LCR  \\
  \hline
  \cite{gutierrez} & 5.8 GHz & Octocopter &  Residential, mountainous & \textendash & RMS, DS, CDF\\
 \hline
\end{tabular}
\end{table}       

\begin{table}[tp] \footnotesize
\renewcommand{\arraystretch}{1.3}

\begin{tabular}{|p{1cm}|p{2cm}|p{1.8cm}|p{1.9cm}|p{1.8cm}|p{4.5cm}|}
  \hline
  \textbf{Ref.} & \textbf{Frequency} & \textbf{UAV} & \textbf{Scenario} & \textbf{Altitude} & \textbf{Channel Statistics} \\
  \hline 
    \cite{Hkung} & 802.11 b/g & Fixed\textendash wing & Farmland & 75 m & AF, DG \\
   \hline
   \cite{qualcomm} &  PCS, AWS, 700 MHz & Quadcopter & Mixed suburban & 122 m & PL, CDF\\
    \hline
   \cite{afonso} & EDGE,HSPA+, LTE & Hexacopter & - & 10-100 m & RTT, J\\
  \hline 
  \cite{teng} & 909 MHz & Quadcopter & Open field, mock village & 40-60 m & PL, PES\\
 \hline
 \multicolumn{6}{|l|}{AF: correlation function, AFD: average fade duration, BC: coherence bandwidth, CDF: cumulative distribution function,}\\
 \multicolumn{6}{|l|}{DG: diversity gain, DS: Doppler spread, J: jitters, K: Rician factor, LCR: level crossing rate, PAS: power azimuth spectrum,}\\
\multicolumn{6}{|l|}{PDF: probability density function, PDP: power delay profile, PES: power elevation spectrum, PL: path loss, RMS: RMS-}\\
\multicolumn{6}{|l|}{delay spread, RTT: round trip time, SF: shadow fading,  $\mu$, $\xi$: mean and standard deviation of Nakagami m factor}\\
\hline
\end{tabular}
\end{table}   

\section{Empirical Channel Models from Measurement Campaigns}
\label{three}
Channel parameters can change frequently with time and space due to cruising capability of UAVs. Therefore, channel characterization is an essential step to study the impact of fast spatial\textendash temporal variations in the UAV channel and consequently to predict the performance of UAV communications. Thus, many measurements campaigns have been launched to corroborate connections between channel parameters and experimental setups such as the flight altitude, the elevation angle, the separation distance between UAV and ground station and the operating environment to identify and model important factors that undermine the communication performances. Despite of all these efforts, there are no unified answers and conclusions still need to be established by means of reliable channel models. In this section we will review the empirical models that characterize AA and AG propagation channels.\\

\subsection{Air\textendash Air (AA) Channel Characterization}
\label{three-one}
The AA channel characterization is particularly essential in multi\textendash UAV networks and aerial wireless sensor network applications, where the characteristics of the AA channel rely on the UAV altitude and relative velocity etc. In \cite{allred}\textendash\cite{ahmed01}, channel characterization has been performed for aerial wireless sensor networks using IEEE 802.15.4 technology, where the AA channel was shown to have better conditions than the AG channel in terms of path loss exponent (PLE). In \cite{allred}, empirical study has been conducted using micro aerial vehicles to characterize the impact of distance and antenna orientation on the received signal strength in the AA channel. The authors have performed linear regression on the samples of received signal and computed PLE for  AA, AG and ground\textendash to\textendash ground (GG) wireless channels. They found that the GG channel performed poorly with PLE of 3.57. PLE for AA and AG channels were estimated to be 1.92 and 2.13, respectively. Similarly, in \cite{shaw}, the received signal strength for AA and AG channels decreases with the separation distance, but at a marginally reduced rate. The path loss exponent was estimated from the log\textendash distance propagation model and PLE to be 0.93 and 1.50 for AA and AG propagation, respectively. On the other hand, the authors of \cite{ahmed01} have observed that the received signal strength for AG, AA and ground\textendash to\textendash air (GA) propagation improves with extended UAV altitude and deteriorates as UAV distance increases. They observed that the AA channel followed free space path loss with PLE of 2.05. Whereas, the presence of gray zones of communication leads to asymmetry in AG and GA channels with PLE of 2.32 and 2.51, respectively. \\

Aerial link characterization has been conducted in \cite{goddemeier02} and \cite{yanmaz03} using IEEE 802.11 radio. In \cite{goddemeier02}, the impact  of UAV altitude on the AA propagation was investigated for large\textendash scale variations and small\textendash scale fading distribution. In this study, path loss was determined by the Friis equation with PLE of 2.6 and fading channel distribution fits with height\textendash dependent Rician factor $K$. In \cite{yanmaz03}, log\textendash distance model was used to analyze the path loss for vertical and horizontal distances. In this work, minimum mean square error (MMSE) method was utilized to compute the PLE of 2.03 and 2.01 for AA and AG channels, respectively.\\
 
The AA channel characterization highlights that the propagation conditions are generally determined by the vertical and horizontal distances between the multiple airborne UAVs in LOS condition. However, significant attenuation occurs for the characterization of aerial link beyond the LOS condition to maintain large communication range. Also it would be useful to study the consequences of the Doppler frequency shift as the multiple UAVs cruises with higher velocities. Large\textendash scale fading statistics of the AA channel are summarized in Table II. \\     

\begin{table}[tp] \footnotesize
	\renewcommand{\arraystretch}{1.3}
	\centering
	\caption{Large\textendash scale fading statistics for AA Channel}
	\label{AA fading}
	\begin{tabular}{|p{1.5cm}|p{12cm}|}
		\hline
		\textbf{Ref.} & \textbf{PL model} \\
		\hline
		\cite{allred}-\cite{ahmed01} & $PL(dB)= 10\alpha\log_{10}(d)$,\newline $\alpha$= 1.922,\cite{allred}, $\alpha$= 0.93\cite{shaw}, $\alpha$= 2.05\cite{ahmed01}\\
		\hline
		\cite{goddemeier02} & $RSS(dB)= P_{t}+G_{UAV1}+G_{UAV2}+10\log_{10}(\frac{\lambda}{4\pi d})^{\alpha}$,\newline $P_{t}$= 20 dBm, $G_{UAV1}=G_{UAV2}$= 5 dBi, $\alpha$= 2.6, $f_{c}$=2.4 GHz\\
		\hline
		\cite{yanmaz03} & $PL(dB)=PL(d_{0})+10\alpha\log_{10}(\frac{d}{d_{0}})$,\newline $d=\sqrt{d_{h}^2+d_{v}^2}$, $PL(d_{0})$= 46.4 dB, $\alpha$=2.03, $d_{h}\in\{0,...,100 m\}$, $d_{v}$= 50 m, $d_{0}$= 1 m\\
		\hline
		\multicolumn{2}{|l|}{$\alpha$: path loss exponent, $RSS$: received signal strength, $d$: separation distance, $d_{0}$: reference distance, $d_{h}$: horizontal-}\\
		\multicolumn{2}{|l|}{distance, $d_{v}$: vertical distance, $G_{UAV}$: UAV antenna gain}\\
		\hline
	\end{tabular} 
\end{table}    

\subsection {Air\textendash Ground (AG) Channel Characterization}
\label{three-two}

\subsubsection{Large\textendash Scale Fading Statistics}
\label{three-two-one}        
 Most of the AG channel measurements focus on the large\textendash scale statistics such as path loss exponent and shadow fading. For an urban environment in \cite{Simunek02}, the measured results exhibited that the path loss follows a distance\textendash independent trend and is significantly affected by the low elevation angle. The excess path loss model is developed by extending terrestrial macro cell models that includes the reflection and diffraction losses caused by the surrounding buildings and incorporated by the knife\textendash edge diffraction theory. For a suburban environment in \cite{Cai}, the impact of UAV altitude and distance on path loss was analyzed. For UAV altitude, simplified dual slope path loss model was considered and found that PLE is negative below the breakpoint altitude because of partially cleared first Fresnel zone, whereas when the UAV altitude increases above the breakpoint level path loss is roughly similar to the free space propagation due to sufficiently cleared first Fresnel zone. For horizontal UAV distance, path loss analyzed with log\textendash distance model. Also, in \cite{khawaja}, the effect of UAV altitude and the optimal placement of ground receiver on path loss was stochastically modeled for both static and mobile UAV in an open field and suburban scenario, while considering foliage losses and Doppler frequency shift. In addition, shadow fading is modeled with zero\textendash mean Gaussian distribution and analyzed with PDF. Another empirical study was conducted in \cite{frew}, to evaluate the influence of distance on path loss attenuation and found degraded performance of the AG channel due to detrimental effect of interference from the other 802.11 devices operated in the surrounding test area. Moreover, in \cite{goddemeier01}, received signal strength declined with the distance and followed the Friis channel model. In \cite{yanmaz01}, the AG propagation channel in the single\textendash hop UAV system followed log\textendash distance model, where higher throughputs were attained over longer distance. \\
 
For an open field and campus environment in \cite{yanmaz02}, path loss was evaluated with the free space path loss model. In \cite{cheng} and \cite{hague}, PLE was estimated using linear regression. In \cite{goddemeier03}, distance and frequency independent empirical path loss model was proposed for urban and rural terrains, where the altitude of aerial mobile station was accounted as the key modeling parameter. In contrast, the empirical propagation model in \cite{tavares} suggested that path loss model is dependent on the distance in 3D plane and the operating frequency. In this case, other modeling parameters such as the UAV altitude and the tilt angle of base station sector antenna were also considered. The altitude dependent path loss model was proposed in \cite{amorim}, where path loss and shadow fading were decreased as the UAV altitude increased from 15 to 120 m and at about 100 m the propagation condition matched to that of free space. In \cite{hourani}, the angle\textendash dependent AG propagation channel model was presented, which encompasses excess path loss attenuation and shadow fading model. In this work, the model parameters are dependent on the angle between cellular base station and airborne UAV. The analytical path loss model was used in \cite{qualcomm} to evaluate the performance of LTE network with UAV platform, where most of the path loss samples computed by measurements were lumped between the reference PLE of 2.0 and 4.0. In \cite{teng}, the combinational model was developed to determine the low coverage zones in the cellular\textendash connected UAV network. This model identified causes such as two\textendash ray ground reflections, diffraction losses and nulls in antenna radiation pattern as the predominant factors for path loss.\\

Path loss and shadow fading statistics for the AG propagation channel presented in this section demonstrated that the UAV flight dynamics, such as the altitude and distance from the ground level, are the dominant contributors for the large\textendash scale fading. Therefore, the development of realistic UAV propagation model requires these parameters to be considered in 3D coordinates. Also, considerable attention is needed for characterizing antenna design and orientations, as this will further improve the UAV communications. In Table III, we summarize the large\textendash scale fading statistics for the AG channel.\\

\begin{table}[tp] \footnotesize
	\renewcommand{\arraystretch}{1.3}
	\caption{Large\textendash scale fading statistics for AG Channel}
	\label{AG fading}
	\begin{tabular}{|p{1.5cm}|p{14cm}|}
		\hline
		\textbf{Ref.} & \textbf{PL model} \\
		\hline
		\cite{Simunek02}& $PL(dB)=-10\log_{10}[\frac{0.05\lambda}{2h^2}(d_{2d}+r_{b}^2d_{2r})]-20\log_{10}(1-e^\varrho)^2$,\newline $\varrho= -0.6038\times0.109^v$, $v\approx h\sqrt{\frac{2}{\lambda d_{2}}}$, $h$= obstruction height, $d_{2}$= distance between receiver and obstruction, $d_{2d}$= direct\textendash ray distance between receiver and obstruction, $d_{2r}$= reflected\textendash ray distance between receiver and obstruction, $r_{b}$= reflection coefficient \\
		\hline
		\cite{Cai} & \underline{Vertical}: {\[
			PL(dB)=
			\begin{cases}
			PL(d_{0}) + 10\alpha_{1}(\log_{10}\frac{d}{d_{0}})&\text{if }d<d_{B}\\
			PL(d_{0}) + 10\alpha_{1}(\log_{10}\frac{d}{d_{0}})+10\alpha_{2}(\log_{10}(\dfrac{d}{d_{B}}))&\text{if }d\geq d_{B},
			\end{cases} 
			\]}\newline $(\alpha_{1}, \sigma_{1}dB)$= (0.74, 1.23), $(\alpha_{2}, \sigma_{2}dB)$= (2.29, 2.15), $d_{B}$= 9 m \\
		&\underline{Horizontal}: $PL(dB)=PL(d_{0}) + 10\alpha(\log_{10}\frac{d}{d_{0}})$,\newline for 20 m: $(\alpha, \sigma dB, PL(d_{0})dB)$=(0.93, 5.5, 77.9), for 30 m: $(\alpha, \sigma dB, PL(d_{0})dB)$=(1.01, 3.9, 74.6) \\
		\hline
		\cite{khawaja} & \underline{Static UAV}: $PL(dB) = PL(d_{0}) + 10\alpha(\log_{10}\frac{d}{d_{0}}) - \log_{10}\frac{\bigtriangleup h}{h_{opt}} + C_{p} + \zeta$,\newline $\bigtriangleup h = \lvert h_{g} - h_{opt} \rvert$, $\bigtriangleup f = (\frac{\bigtriangleup v}{c})f_{c}$, $ \zeta \sim N(0,\sigma^{2})$, $C_{p}$=0 dB, $d$= 5.6 m to 16.5 m, $h_{g}$= (1.5 m, 7 cm),  $(\alpha, \sigma dB, PL(d_{0})dB) $=2.6471, 3.37, 34.905 (open, 0 km/h), $(\alpha, \sigma dB, PL(d_{0})dB)$=2.7601, 4.8739, 30.4459(suburban, 0 km/h),\\
		&  \underline{Mobile UAV}: $PL(dB) = PL(d_{0}) + 10\alpha(\log_{10}\frac{d}{d_{0}}) - \log_{10}\frac{\bigtriangleup h}{h_{opt}} + C_{p} + 10\textit{x}\log_{10}(\frac{f_{c}+\bigtriangleup f}{f_{c}}) + \zeta $,\newline , $(\alpha, \sigma dB, PL(d_{0})dB)$=2.6533, 4.02, 34.906(open, 32 km/h), $(\alpha, \sigma dB, PL(d_{0})dB)$=2.8350, 5.3, 30.446(suburban, 32 km/h), $x$= frequency dependent path loss factor and negligible at small velocities  \\
		\hline
		\cite{frew} & $RSS(dBm)=-95+10\log_{10}(K_{0_{}}.d^{-\alpha})$, $\alpha$=2.34, $K_{0}=3.6\times10^{-1}$ \\
		\hline 
		\cite{goddemeier01} & $RSS(dB)= P_{t}+G+10\log_{10}(\frac{\lambda}{4\pi d})^{\alpha}$, $P_{t}$= 20 dBm, $G$= 1dB, $f_{c}$=2.4 GHz, $\alpha$=2.3 \\
		\hline
		\cite{yanmaz01} & $PL(dB)= 10\alpha\log_{10}(d)$, $\alpha$ $\approx$2 for beyond 100 m distance \\
		\hline
		\cite{allred}-\cite{ahmed01} & $PL(dB)= 10\alpha\log_{10}(d)$, \newline $\alpha$=2.132 (AG), 3.57 (GG)\cite{allred}, 1.50 (AG) \cite{shaw}, 2.32 (AG), 2.51 (GA), 3.1 (GG)\cite{ahmed01}  \\
		\hline
		\cite{yanmaz03} & $PL(dB)=PL(d_{0})+10\alpha\log_{10}(\frac{d}{d_{0}})$,\newline $d=\sqrt{d_{h}^2+d_{v}^2}$, $PL(d_{0})$= 46.4 dB, $\alpha$=2.01 (AG), $d_{h}\in\{0,...,100 m\}$, $d_{v}$= 50 m, $d_{0}$= 1 m\\
		\hline
		\cite{yanmaz02} & $RSS(dBm)=P_{rx}(d_{0})-10\alpha\log_{10}(\frac{d}{d_{0}})$, $P_{rx}$: receive power at reference distance, $\alpha$=2.2 (open), 2.5-2.6(campus) \\
		\hline
\end{tabular} 
\end{table}       

\begin{table}[tp] \footnotesize
	\renewcommand{\arraystretch}{1.3}
	\begin{tabular}{|p{1.5cm}|p{14cm}|}
		\hline
		\textbf{Ref.} & \textbf{PL model} \\
		\hline
		\cite{cheng},\cite{hague} & $RSS(dBm)=A-10\alpha log_{10}(d)$, $\alpha$=1.80, $A$=-37.5\cite{cheng}, $\alpha$=1.04 , $A$=-55.12\cite{hague}  \\
		\hline
		\cite{goddemeier03} & \underline{Urban:} $PL(dBm) = 89.5357 + (\frac{h_{UAV}^{3}}{10000} + 0.0108h_{UAV}^{2} + 0.8588h_{UAV})$\newline \underline{Rural}: $PL(dBm) = 78.2186 - 0.0013h_{UAV}^{2} - 0.0052h_{UAV}$, \newline $h_{UAV}\in\{0,...,500 m\}$ \\
		\hline 
		\cite{tavares} & $PL(dB) = 20\log(\frac{4\pi d_{0}}{\lambda}) +X_{dis}+X_{freq}+X_{hei}+X_{ang}$, \newline $X_{dis}, X_{freq}, X_{hei}, X_{ang}$: model parameters in 3D plane \\
		\hline
		\cite{amorim} & $PL(dB) = \alpha(h_{UAV})10\log_{10}(d) + \beta(h_{UAV}) + \zeta $, \newline $ \zeta \sim N(0,\sigma(h_{UAV}))$, $\alpha(h_{UAV})$= 2.9-2.0 (15-100 m), $\beta(h_{UAV})dB$= -1.3-35.3 (15-100 m), $\sigma(h_{UAV})dB$= 7.7-3.4 (15-100 m)  \\
		\hline
		\cite{hourani} & $PL(dB) = \alpha10\log_{10}(d) + A(\phi - \phi_{0})\exp(-\frac{\phi - \phi_{0}}{B}) + \eta_{0} + \zeta $ \newline $\zeta \sim N(0,a\phi + \sigma_{0})$, $\alpha$= 3.04, $A$= -23.29, $B$= 4.14, $\phi_{0}$= -3.61, $\eta_{0}$= 20.70, $a$= -0.41, $\sigma_{0}$= 5.86  \\
		\hline 
		\cite{qualcomm} & $PL(dB)=P_{tx}-10\log_{10}(12.BW)-RSRP+G_{UAV}+G_{BS}$ \newline $P_{tx}$= maximum transmit power, $BW$= transmission bandwidth, $RSRP$= measured reference signal received power, $G_{UAV}$= gain of UAV antenna, $G_{BS}$= gain of base station antenna  \\
		\hline
		\cite{teng} &$PL(dB)=-20\log_{10}|\nu|+40\log_{10}(d)-10\log_{10}(h_{BS}^2h_{UAV}^{2})$,\newline $\nu$: Kirchoff diffraction parameter \\
		\hline
		\multicolumn{2}{|l|}{$\alpha$= path loss exponent, $RSS$= received signal strength, $d$= separation distance, $d_{0}$= reference distance, $d_{B}$= breakpoint-}\\
		\multicolumn{2}{|l|}{distance, $C_{P}$= foliage loss, $\sigma$= standard deviation, $h_{g}$= height from ground level, $h_{opt}$= optimal height from ground level,}\\
		\multicolumn{2}{|l|}{$f_{c}$= carrier frequency, $h_{UAV}$= UAV altitude, $h_{BS}$ base station altitude, $\bigtriangleup f$= Doppler shift, $K_{0}$= transmission gain,}\\
		\multicolumn{2}{|l|}{ $G$= antenna gain, $A$= \textit{y}\textendash intercept, $\lambda$= wavelength}\\
		\hline
	\end{tabular} 
\end{table}         

\subsubsection{Small\textendash Scale Fading Statistics}
\label{three-two-two}
Multipath propagation and small\textendash scale channel characterization is important to study the impact of the fading channel behavior on UAV communications. In \cite{Simunek01}, narrowband fading characteristics were analyzed for the low elevation propagation channel and time\textendash series generator was developed to capture the channel effects. Hence, the variations of the received signal voltage ($y$) were statistically analyzed with the PDF of combination of Rician and Log\textendash normal distribution.
\begin{eqnarray}
f(y) = \frac{y}{\sigma^2 \sqrt{2\pi \varSigma_{A}^2}}\int_{a=0}^\infty \frac{1}{a}e^{\frac{-(20\log a-M_{A})^2}{2\varSigma_{A}^2}}
e^{\frac{-(y^{2}+a^{2})}{2\sigma^{2}}}I_{0}(\frac{ya}{\sigma^{2}})da,
\end{eqnarray}
where $M_{A}$ and $\varSigma$ are respectively, the mean and standard deviation of the Gaussian distribution for the magnitude of the LOS signal, $a$ and $\sigma$, respectively, denotes the magnitude of the LOS and the diffuse multipath components of Ricean distribution, respectively, and $I_{0}(.)$ is the zeroth\textendash order modified Bessel function.\\

Rician and Log\textendash normal compositional models are preferred for the land mobile satellite systems and referred as Loo model \cite{Chini}. The fading channel variations for these models are studied with the first order statistics, such as cumulative distribution function (CDF) or probability density function (PDF), and second order statistics such as, average fade duration (AFD) or level crossing rate (LCR). Autocorrelation function and power spectral density was used for the proposed time series generator to reproduce channel dynamics.\\

Rician distribution was used in \cite{Cai} and \cite{goddemeier02} to analyze the time varying effects in the AG propagation channel. In this case, the dominant LOS and diffuse multipath components were fitted using the PDF of Rician distribution given by
\begin{eqnarray}
f(y) = \frac{y}{\sigma_{2}}e^{\frac{-(y^{2}+a^{2})}{2\sigma^{2}}}I_{0}(\frac{ya}{\sigma^{2}}),
\end{eqnarray}
where $y\geq0$, $a$ and $\sigma$ denotes the magnitude of the LOS and the diffuse multipath components, respectively. Also, the Rician parameter $K$ is defined as:
\begin{eqnarray}
K=\frac{a^{2}}{2\sigma^{2}}.
\end{eqnarray}

In \cite{Cai}, small\textendash scale fading parameters were approximated by the SAGE algorithm and analyzed to study the behavior of multipath components at different altitude levels. Also, in \cite{goddemeier02}, multipath propagation effects caused by the ground reflection were analyzed by the height\textendash dependent Rician parameter $K$.

In \cite{khawaja} and \cite{yanmaz03}, the Nakagami m distribution was used with PDF given by
\begin{eqnarray}
f(y;m,\varOmega) = \frac{2m^{m}}{\Gamma(m)\varOmega^{m}}(y^{2m-1})e^{(\frac{-my^{2}}{\varOmega})},
\end{eqnarray}
where $\Gamma(m)$ is the Gama function, $m$ and $\varOmega$ are the Nakagami shape and spread controlling parameters, respectively,
\begin{eqnarray}
m = \frac{E^{2}[X^{2}]}{Var[X^{2}]}\\
\varOmega = E[X^{2}].
\end{eqnarray}

In \cite{khawaja}, the magnitude of individual multipath components was collected for different time delay bins from the multiple channel impulse responses and modeled by the Nakagami m distribution, where the mean ($\eta$) and standard deviation ($\xi$) of the $m$ parameter were evaluated empirically. As a result, mean is found to be small for both open and suburban areas under the influence of tree canopy and large variance is observed due to thick suburban scattering. In addition, multipath channel characteristics, such as time of arrival, was modeled as a Poisson process and analyzed by the CDF. Also, time dispersive parameters were estimated by Clean's algorithm. In this case, frequency dispersive parameters were not computed due to the low velocity of UAV. In \cite{yanmaz03}, the CDF analysis was performed to compare the theoretical Rayleigh and Nakagami distribution. It was observed that as the shape parameter $m$ is always greater than 1. Thus, Nakagami m fading channel has the best fit.\\

Time and frequency dispersive parameters were computed in \cite{gutierrez} for residential and mountainous desert terrain, where cross ambiguity function (referred as scattering function) was opted to estimate channel parameters from the measured channel impulse response. In this work, the mean and variance of the power delay profile were compared by CDF and it was found that for the mountainous desert scenario, the median RMS delay spread and the Doppler frequency spread are roughly 0.06$\mu s$ and 28.96 Hz, respectively. For the residential area, the measured median RMS delay spread and the Doppler frequency spread are approximately 0.03$\mu s$ and 28.06 Hz. The RMS delay spread attained in the desert terrain is larger due to the rough mountainous scatters along the flight path than those measured in the residential area.\\

Channel characterization for small\textendash scale fading effects mostly addresses temporal variations and scant efforts have been put to study spatial variation. Also, for most AG propagation cases reported, the Nakagami and Rician distributions seem to effectively analyze the fading channel statistics.\\

\section{Analytical  Channel Models}
\label{four}
Analytical channel models are useful for characterizing the propagation behavior under certain assumptions and parameters. They can predict the performance of communication systems. For the land mobile satellite systems, channel behavior can be analyzed using the multi\textendash state Markov chain model \cite{Panagopoulos},\cite{Chini}. For the terrestrial cellular systems, there are three main modeling approaches: deterministic, stochastic and geometry\textendash based stochastic approach. In the \textit{deterministic} technique, environmental\textendash specific parameters are utilized to model the propagation channel. In the \textit{stochastic} approach, the propagation characteristics are realized by the channel statistics without requiring the specific location information. In the \textit{geometry\textendash based stochastic} approach, random scatters are assumed in the environment to obtain the spatial\textendash temporal statistics in a stochastic manner by applying the deterministic or ray\textendash tracing model. These models can be used for UAV channels. In this section, we will categorize the UAV channel modeling approaches reported in the literature and presents some analytical expressions of them.\\ 

\subsection{Deterministic}
\label{four-one}
In deterministic channel models, environmental clutters are placed in the certain layouts. This approach assumes large dimensions of the environmental objects in comparison with the wavelength, thus not compensating the diffuse scattering. The accuracy of these channel models depends on the environment\textendash specific database which consists of the information related to the terrain topography, the electrical parameters of buildings and other obstruction materials. Deterministic models can also be realized by the ray\textendash tracing simulation software, which can depict the realistic behavior of the EM wave propagation and simulate power loss and shadowing effects.\\ 

In \cite{Daniel}, 3D ray\textendash tracing was performed to characterize the AG propagation in the suburban environment for the channel between cellular base station and airborne UAV. Also, the well\textendash known macro cell terrestrial channel models were tested with the boundary conditions in order to determine the applicability of low altitude aerial platforms for providing cellular coverage. In \cite{Feng01} and \cite{Feng02}, analytical propagation models have been studied for the AG channel characterization in an urban environment for the frequency ranging from 200 MHz to 5 GHz and the aerial altitude from 100 to 2000 m. In \cite{Feng01}, the average path loss and shadowing statistics were examined as the function of elevation angle and the aerial altitude through 3D ray\textendash tracing simulations. The authors have provided analytical path loss expressions. Also, the shadow fading was fitted with the log\textendash normal distribution with the standard deviation dependent on the elevation angle. In addition, propagation conditions were determined from the simulation results and categorized as, LOS, NLOS (non-LOS) and OLOS (obstructed LOS) channels. The work in \cite{Feng02} utilized knife\textendash edge diffraction theory to model the LOS probability, which considered the statistical parameters to account for height, size and coverage area of buildings in the simulation environment.\\

In \cite{hourani01}\textendash \cite{hourani03}, environmental topography was realized with the statistical parameters recommended by the International Telecommunication Union (ITU\textendash R). In \cite{hourani01}, a generic path loss model in low altitude platform was proposed, where the channel model parameters were estimated by the 3D ray\textendash tracing at 700 MHz, 2000 MHz and 5800 MHz. In this work, the AG channel conditions favoring LOS and NLOS propagations were grouped distinctly and analyzed with the group occurrence probability as the conditional PDF. Simulation results demonstrated that the impact of elevation angle was significant on the excess path loss.

In \cite{hourani02}, the closed\textendash form expressions was formulated for predicting the coverage footprint from the aerial platform in terms of the maximum cell radius and the optimal altitude. In this study, the free space path loss model was extended with the excessive attenuation factor and corresponds to LOS and NLOS propagation conditions. This was extended in \cite{hourani03} to provide the analytical framework for optimization of the average radio coverage probability and the maximum transmission rate to achieve the required quality of service. Some of the deterministic UAV channel models are reported in Table IV  \\  

\begin{table}[tp] \footnotesize
	\renewcommand{\arraystretch}{1.3}
	\caption{Deterministic Models}
	\label{deterministic modeling approaches}
	\centering
	\begin{tabular}{|p{0.5cm}|p{7cm}|p{7cm}|}
		\hline
		\textbf{Ref.} & \textbf{Analytical Model} & \textbf{Parameters}\\
		\hline
		\cite{Feng01} & \underline{Path loss}: \[PL(dB)=
		\begin{cases}
		-0.58 + 0.549\textit{e}^{\frac{(90 - \phi)}{24}}&\text{LOS channel}\\
		\eta_{0} - \eta_{1}\textit{e}^{-\frac{(90 - \phi)}{\nu}}&\text{NLOS channel}\\
		\iota_{0} - \iota_{1}\textit{e}^{\frac{(90 - \phi)}{\omega}}&\text{OLOS channel}\\
		\end{cases}\] \newline  \underline{Shadow fading}: $\sigma(dB)= \rho(90 - \theta)^{\gamma}$ & $\phi$: elevation angle, for 200 MHz: ($\eta_{0},\eta_{1},\nu$)= (9.08, 6.40, 12.01), ($\iota_{0}, \iota_{1}, \omega$)= (2.11, 0.41, 22.07), for 5000 MHz: ($\eta_{0},\eta_{1},\nu$)= (20.43, 14.60, 10.50), ($\iota_{0}, \iota_{1}, \omega$)= (6.23, 0.4787, 22.65), LOS channel at 200 MHz and 100 m altitude: ($\rho,\gamma$)= (0.0143, 0.9941), NLOS channel at 200 MHz \& 5000 MHz: ($\rho,\gamma$)= (0.7489, 0.4638) \& (2.7940, 0.2259), OLOS channel at 200 MHz \& 5000 MHz: ($\rho,\gamma$)= (0.3334, 0.3967) \& (0.8937, 0.3713) \\
		\hline  
		\cite{Feng02} & \underline{LOS probability in a street}: \[P_{LOS}=
		\begin{cases}
		1-\frac{S_{c} \sin\vartheta}{W_{s}},&\text{$0<S_{c}<W_{s}/ \sin\vartheta$}\\
		0, &\text{$S_{c}>W_{s}/ \sin\vartheta$}\\
		\end{cases}\] & $\phi$: elevation angle, $\vartheta$: street angle, $\vartheta_{c}$: critical street angle, $W_{s}$: street width, $S_{c}$: critical distance between ground station to adjacent buildings, $H$= building height, $h_{g}$: ground station height, $W_{e}$: estimated street width, $\lambda$: wavelength, $W_{s}$= 15m, $\vartheta$= $90^{\circ}$, $\Delta H$=$H-h_{g}$, $H$= 11.71m, $h_{g}$= 15m, $W_{e}$=44.2m     \\
		& \underline{LOS probability in an area}: & $S_{c}>\Delta H\cot\phi+\frac{0.16\lambda\cos\phi+\sqrt{(0.16\lambda\cos\phi)^2+0.32\lambda\Delta H\sin\phi}}{\sin^2\phi}$,\\
		& $P_{LOS}= \frac{2}{\pi}[\vartheta_{c}-\frac{S_{c}(1-\cos\vartheta_{c})}{W_{e}}]$ &  \[\sin\vartheta_{c}=\begin{cases}
		\frac{W_{e}}{S_{c}},&\text{$W_{e}\leq S_{c}$}\\
		0, &\text{otherwise}\\
		\end{cases}\] \\
		\hline
		\cite{hourani01} & \underline{Path loss}:\newline $PL(dB)= 20\log(\frac{\bigtriangleup h}{\sin\phi})+ 20\log(f_{MHz}) - 27.55$
		& $h_{UAV}$: UAV altitude, $\bigtriangleup h= h_{UAV}-h_{g}$, $h_{UAV}$= 200m, $h_{g}$= 1.5 m, $f_{MHz}$= 700, 2000, 5800\\
		& \underline{LOS probability}: $P_{LOS}= a(\phi-\phi_{0})^b $, & $\phi_{0}= 15^{\circ}$, For 700 MHz: suburban($a= 0.77, b= 0.05$), urban($a= 0.63, b= 0.09$), dense urban($a= 0.37, b= 0.21$), high\textendash rise urban($a= 0.06, b= 0.58$)  \\
		\hline
		\cite{hourani02} & \underline{Path loss}: $PL(dB)$= \[\begin{cases}
		20\log(\frac{4\pi f_{c}d}{c})+\varepsilon_{LOS} &\text{LOS channel}\\
		20\log(\frac{4\pi f_{c}d}{c})+\varepsilon_{NLOS} &\text{NLOS channel}\\
		\end{cases}\] & $d=\sqrt{h_{UAV^2+r^2}}$, $r=$ cell radius, $f_{c}= 2000 MHz$: suburban($\varepsilon_{LOS}= 0.1, \varepsilon_{NLOS}= 21$), urban($\varepsilon_{LOS}= 1.0, \varepsilon_{NLOS}= 20$), dense urban($\varepsilon_{LOS}= 1.6, \varepsilon_{NLOS}= 23$), high\textendash rise urban($\varepsilon_{LOS}= 2.3,\varepsilon_{NLOS}= 34$) \\
		\hline
	\end{tabular}
\end{table}
\subsection{Stochastic Channel Model}
\label{four-two}
For the UAV communication systems, stochastic based channel models can be designed using the tapped delay line (TDL) system with different numbers of taps, each of which can accommodate fading statistics of the multipath components derived from the channel impulse response. The accuracy of these model depends on the estimation of stationary interval in the non\textendash stationary UAV channel.\\

In \cite{Matolak03}\textendash\cite{Matolak05}, wideband stochastic channel models were proposed from the data collected in different environments, using the estimated stationary interval of 15 m at the C band. For the over water settings in \cite{Matolak03}, the AG channel employed the TDL model to characterize the two\textendash ray propagation plus an intermittent multipath component as the third ray. In this work, the authors have argued that the statistics for LOS and reflected components can be analyzed by either curved earth two ray (CE2R) or flat earth two ray (FE2R) model. The probability of the existence of the intermittent multipath component was estimated by the exponential distribution as a function of link distance. The TDL model with nine taps have been proposed for the mountainous terrain \cite{Matolak04} and an urban environment \cite{Matolak05}. In both of these studies, seven intermittent multipath components were considered and the probability of existence, excess delay and duration of intermittent multipath components were modeled as the linear function of link range. 

In \cite{elnoubi} and \cite{zaman}, stochastic model was developed with the narrowband assumption to characterize the aeronautical AG channel. In \cite{elnoubi}, the stochastic model was designed for characterizing the AG propagation in terms of transmission coefficients assuming that the quadrature components reflected from the ground surface can be modeled as a zero\textendash mean Gaussian process. Also, Doppler spectrum analysis was performed for the diffuse multipath components. In \cite{zaman}, the proposed model was developed with the TDL system having both LOS and NLOS taps, where the amplitude attenuation and the multipath delay of NLOS components were assumed to be Rayleigh distributed and Gaussian random process, respectively, while the phase shift was uniformly distributed. Further, the Doppler frequency shift was characterized as the time varying random process and the channel stationarity interval was not computed, but the fading statistics were assumed to be constant for the random time duration. In Table V, we have presented the channel response from the TDL models reported in this paper. \\
\begin{table} [tp] \footnotesize
	\renewcommand{\arraystretch}{1.3}
	\caption{TDL Models}
	\label{Stochastic modeling approaches}
	\centering
	\begin{tabular}{|p{1cm}|p{14cm}|}
		\hline
		\textbf{Ref.} & \textbf{TDL model}\\
		\hline
		\cite{Matolak03} & $h(\tau,t)= h_{2-ray}(\tau,t)+w_{3}(t)A_{3}(t)\exp(-j\varphi_{3}(t))\delta(\tau-\tau_{3}(t))$,\newline $h_{2-ray}$ denotes FE2R or CE2R model, $w_{3}(t)\in\{0,1\}$ represents presence/absence of third\textendash ray and modeled as $p(d)=ae^{bd}$, $A_{3}$ is the amplitude of third\textendash ray and modeled by the Gaussian distribution, $\varphi_{3}\in\{0,2\pi\}$ is the uniformly randomly distributed phase of third\textendash ray, $\tau_{3}$ is the excess delay of third\textendash ray and modeled as $p(\tau_{3})= \frac{1}{\mu}e^{-(\tau_{3}-100/\mu)}$, $(a,b)$=(0.17,-0.25) over sea water and (0.03,-0.15) over freshwater, $\mu$= 17 ns, $6 ns\leq\tau_{3}\leq 7 ns$, $d$ is the link distance \\
		\hline
		\cite{Matolak04}, \cite{Matolak05} & $h(\tau,t) = A_{1}(t)\delta(\tau-\tau_{1}(t))+A_{2}(t)\exp(-j\varphi_{2}(t))\delta(\tau-\tau_{2}(t))  +\sum_{L=3}^9w_{L}(t)A_{L}(t)\exp(-j\varphi_{L}(t))\delta(\tau-\tau_{L}(t))$ \newline $A$, $\varphi$ and $\tau$ denotes amplitude, phase and excess delay, respectively, subscripts 1, 2 and L represents LOS, reflected and $L^{th}$ intermittent multipath components, respectively, variations of $w_{L}$ and $\tau_{L}$ are modeled as a linear function of link range , $\varphi_{L}\in\{0,2\pi\}$ is the the uniformly randomly distributed phase, $10\log(\frac{A_{L}^2}{A_{1}^2})$ represents relative power of intermittent components and follows a Gaussian distribution\\
		\hline
		\cite{zaman} & $y(t)= A_{1}(t)\cos[2\pi\{f_{c}+\bigtriangleup f\}(t-\tau_{1}(t))]+\sum_{L=2}^N A_{L}(t)\cos[2\pi\{f_{c}+\bigtriangleup f\}(t-\tau_{L}(t))+\varphi_{L}(t)]+n(t)$  \newline $A_{1}$ is the amplitude of LOS path, $A_{L}$ represents amplitude of NLOS paths and assumed as Rayleigh random process, $\varphi_{L}\in\{-\pi,\pi\}$ is the phase shift of NLOS paths and modeled as uniform random process, $\bigtriangleup f$ denotes Doppler frequency shift and modeled as time\textendash variant random process, $\tau_{1}$= 25$\mu$s, $\tau_{L}$ is excess delay of NLOS components and modeled as the Gaussian random process with mean and standard deviation of 30$\mu$s and 5$\mu$s, respectively, $n(t)$ is white Gaussian noise      \\
		\hline   
	\end{tabular}
\end{table}
\subsection{Geometry\textendash based Stochastic Channel Model}
\label{four-three}
Geometry\textendash based stochastic modeling approach obtains the spatial\textendash temporal channel characteristics with stochastic output in a 3D geometric simulated environment. The accuracy of this model is dependent of the physical environment surrounded with scatters following certain probability distributions. The geometric based channel model for the analysis and simulation of the AG radio communication was proposed in \cite{newhall01}. It characterized the multipath propagation in a cluttered environment around the ground station confined within a virtual 3D ellipsoidal geometry to analytically evaluate delay, gain, phase and angle of arrival (AOA) of individual multipath components. The path loss can be determined using the log\textendash distance model between the airborne platform and the clutters. Therefore, the proposed model is equally applicable to determine both narrowband and wideband channel statistics and well suited for designing antenna diversity system and antenna arrays. This was extended in \cite{wentz} to theoretically estimate the MIMO performance for the low altitude AG channel and also characterize the propagation loss for LOS and multipath components using the log\textendash distance path loss model with the log\textendash normal shadow fading. In this model, the small\textendash scale spatial fading was modeled by the Ricean distribution to analyze the scattering of the multipath components. Furthermore, the probability of error was simulated for SISO system and compared with a $2\times2$ space time block coding and a $2\times2$ spatial multiplexing gain using maximum likelihood detection. In \cite{ibrahim}, 3D AG propagation model was proposed for the dense scattering environment considering low altitude platform. The model was derived for a direction of arrival and the delay dependent Doppler spectrum with the approximation of linear distribution of the scattering point. In this work, the analytical results were compared with the terrain based digital elevation model simulation results and found that the terrain morphology affects the Doppler\textendash delay spread spectrum. \\

In \cite{gulfam}, a realistic 3D geometric\textendash based stochastic model has been developed for the AG communication between an airborne platform and the base station as an elevated plane. The proposed model considered scattering points as uniformly distributed around the base station. In this study, the spatial characteristics were analyzed with the closed\textendash form analytical expressions. In \cite{zeng}, geometric\textendash based stochastic approach has been utilized for UAV channel modeling to analytically characterize a $2\times2$ MIMO enabled AG propagation in 3D plane. In this case, the model was developed with the assumption that the ground scatters are distributed on the cylindrical surface and scatter free airborne environment. Based on the proposed model, analytical expressions were used to study the impact of elevation angle and direction of UAV movement on the space time correlation function in a non\textendash isotropic environment. Some analytical expressions to determine AOA using geometry\textendash based model are given in Table VI. 
\begin{table} [tp] \footnotesize
	\renewcommand{\arraystretch}{1.3}
	\caption{Geometry\textendash based stochastic model}
	\label{Geometric Stochastic modeling approaches}
	\centering
	\begin{tabular}{|m{1cm}|m{14cm}|}
		\hline
		\textbf{Ref.} & \textbf{Geometry\textendash based stochastic model} \\
		\hline
	\cite{newhall01} & The PDF of AOA as the function of elevation angle $(\phi)$ around the ground receiver: $f(\phi)= \frac{(\frac{x_{a}^2}{x_{a}^2-x_{b}^2})-1}{2\pi\gamma(\frac{x_{a}}{\sqrt{x_{a}^2-x_{b}^2}}-\cos\phi)^2}$, \newline $x_{a}$ and $x_{b}$ are subsequently the major and minor axis of the planar elliptical scattering surface, $\gamma= \frac{x_{a}}{\sqrt(x_{a}^2-x_{b}^2)}[\frac{x_{a}^2}{x_{a}^2-x_{b}^2}-1]^\frac{1}{2}$\\
		\hline
	\cite{gulfam}	&The PDF of AOA seen at the airborne platform: $f(\Psi_{ap},\phi_{ap})= \frac{(l_{ap,max}^{3}-l_{ap,min}^{3})\cos\phi_{ap}}{3V}$, \newline $\Psi_{ap}$ and $\phi_{ap}$ are, respectively, the azimuth and the elevation angle observed from the airborne platform, $l_{ap,max}$ and $l_{ap,min}$ are the distance between the UAV and the farthest and nearest scatter point, respectively.\\
		&The PDF of AOA seen around the elevated ground plane: $f(\Psi_{bs},\phi_{bs})= \frac{(l_{bs,max}^{3}-l_{bs,min}^{3})\cos\phi_{bs}}{3V}$, \newline $\Psi_{bs}$ and $\phi_{bs}$ are, respectively, the azimuth and the elevation angle observed from the base station, $l_{bs,max}$ and $l_{bs,min}$ are the distance between the base station and the farthest and nearest scatterer point, respectively, and $V$ is the volume of the scattering region\\
		\hline
	\cite{zeng}	& The von Mises PDF of AOA as the function of azimuth angle:$f(\Psi) = \frac{e^{k\cos(\Psi-\Psi_{\mu})}}{2\pi I_{0}(k)},-\pi\Psi\leq\pi$, \newline $k$ is a spreading control parameter, $\Psi_{\mu}\in[-\pi,\pi]$ is the mean angle of the distribution of scatterers in a 2D plane, $I_{0}(.)$ is the modified Bessel function of the zeroth\textendash order, $k$=3, $\Psi_{\mu}$=$\pi$\\
		& The cosine PDF of AOA as the function of elevation angle:$f(\phi) =\dfrac{\pi}{4\phi_{m}}\cos(\frac{\pi}{2}\frac{\phi-\phi_{\mu}}{\phi_{m}})$, \newline mean angle $\phi_{\mu}$= $\frac{\pi}{6}$ and variance $\phi_{m}$=$\frac{\pi}{4}$\\
		\hline
	\end{tabular}
\end{table}

\section{Important Issues}
\label{five}

\subsection{Airframe Shadowing}
\label{five-one}
In aeronautical communication, the radio path between aircraft and ground control station may be blocked by aircraft structure, such as wings, fuselage or engine. Also, during flight maneuvering or banking turns, the direct LOS path may severed and thus, induced shadowing. In the context of UAV channel characterization, airframe shadowing is still unexplored, as most of the measurement campaigns pertinent to study this phenomena are initiated with manned aircrafts in high altitude. Therefore, the characterization of airframe shadowing with multi\textendash rotor UAVs in low altitude is an interesting topic. Here, we can only have airframe shadowing characterization with manned aircrafts.\\

In \cite{holzbock}, channel measurements were extracted for the communication link between aircraft and satellite, where shadowing effect was induced by the forced aircraft maneuvering. Characterization of the AG channel in C band was performed in \cite{lee01}, where the propagation signal was obstructed by the right banking turn when the aircraft moving in the circular flight track and led to the weakest received signal. Similarly, strongest signal was detected during the left banking turn. Related measurements were conducted in \cite{lee02}, where CDF analysis of the received signal power during the circular flight track demonstrated that the airframe shadowing can be modeled by the Gaussian distribution. The authors have observed that the shadowing effects were substantial in the circular flight track than in linear flight profile. In \cite{kunisch}, airframe shadowing was reported due to wings and engine of the commercial A320 aircraft, where shadowing statistics were simulated by the finite difference time domain approach. Empirical airframe shadowing model was proposed in \cite{Matolak06}, for analyzing shadowing loss and duration. In this study, aircraft followed oval flight route and found that shadowing statistics were disjoint from the ground environment and link distance.\\

\subsection{Stationary Interval}
\label{five-two}
One of the most important characteristics that distinguish UAV communication from the conventional terrestrial wireless systems is the non\textendash stationarity in UAV channels, when the WSSUS assumption is violated. Therefore, wideband frequency\textendash dispersive channel statistics can possess any significance within the stationary interval of the non\textendash stationary UAV channel. No comprehensive study is available in the literature that addresses channel non\textendash stationarity for the UAV propagation channel in low altitude platform. Therefore, estimation of the stationary interval is a contemporary research topic. Efforts to characterize the AG channel with stationarity interval was performed in \cite{Matolak02}, using manned aircraft at high altitude platform. In this study, stationary interval was computed for the wideband measurements using temporal PDP(power delay profile) correlation coefficient method, whereas, spatial correlation collinearity was considered for narrowband measurements. The estimated stationary interval from both of these methods are approximately 15 m or $250\lambda$ at C band with 50 MHz bandwidth.  

\subsection{Diversity Gain}
\label{five-three}
Diversity techniques are beneficial to enhance the reliability of the communication systems, particularly when deep fades dominate. Diversity possibilities have been mostly exploited in MIMO airborne communications with manned aircrafts. For example, in \cite{Jchen}, a $4\times4$ MIMO enabled OFDM system was used to increase the average throughput by 2 times and the range extension by 1.6 times in comparison to a SISO system. In \cite{Mrice01}, multiple helicopter mounted antennas were utilized to achieve the signal\textendash to\textendash noise (SNR) gain of approximately 13 dB. In \cite{Yjiang}, the spatial multiplexing gain was achieved with a $2\times2$ MIMO configuration and as a consequence the throughput gain was enhanced up to 8 times for most of the flight route.\\

In the context of UAV communications, there are few measurement campaigns on the effect of multiple antenna elements. In \cite{Simunek03} and \cite{Simunek04}, the AG channel characterization was initiated with a $1\times4$ antenna configuration. In this work, carrier\textendash to\textendash noise ratio (CNR) gain was compared for the common combining strategies such as selection, equal\textendash gain and maximal ratio combining (MRC) with different antenna elements and observed that the diversity gain achieved with the MRC method using four antenna elements is approximately 4 dB greater in an urban environment than in a wooded area under similar circumstances.\\

In \cite{Hkung}, the performance of multiple receiver and transmitter nodes was evaluated by the correlation coefficient. In this case, the packet delivery rate was boosted by 25\% on average due to the poor correlation at the multiple receiver nodes in a $1\times4$ configuration and by 37\% with the selection diversity using three transmitters in a $3\times4$ setup. Measurement analysis of a $4\times4$ MIMO channel in \cite{willink} revealed that despite of the sparse multipath environment, poor spatial correlation provides the significant capacity gain due to the planar wavefronts generated by near\textendash field reflections at the ground receiver side. It is argued that more diversity gain could be achieved with the robust airborne platform, constructed from the conventional aircraft materials. In this study, ground spatial characteristics were estimated by the Cosine Hermitian angle.\\  

The diversity gain achieved by multiple antennas are dependent on the number of transmitter and receiver antennas and the operating environment of both the UAV and the ground station. Not enough experimental setups were designed to comprehensively characterize the viability of these systems. Substantial research works are still required to fully recognize the benefits of MIMO technology for both AG and AA propagation.\\  

\section{Future Research Challenges}
\label{six}
In this section, we will discuss some future research challenges for characterizing the UAV channel with measurement campaigns and in the development of realistic UAV channel model: 
\begin{itemize}
	\item To provide seamless coverage with UAV communication in all circumstances, measurement campaigns are required to investigate the UAV channel in dense urban environment and in the metropolitan cities with consent from the local civil aviation regulatory bodies.
	\item USRP platforms can provide more flexibility to initiate UAV measurement campaigns with wideband frequencies and low power consumption. This also allows USRP to test different wireless communication protocols such as multi\textendash carrier and MIMO system to be used in UAV communication.
	\item The AA channel characterization is required to study the consequences of Doppler shift experienced by multiple UAVs cruising with different velocities.  
	\item Airframe shadowing has not received commensurate level of attention for small size rotatory UAVs, Therefore, measurement campaigns are required to study this phenomenon for both AG channel in the single\textendash hop network and AA propagation in the multi\textendash hop network. In addition, ray\textendash tracing can be used to probe the airframe shadowing, as CAD tools are capable of incorporating UAV shape, metallic properties and different maneuvering positions.
	\item Estimation of the stationary interval is of paramount importance for characterizing the AG channel regarding wideband frequency\textendash selective parameters. Therefore, it would be interesting to estimate channel parameters with spectral divergence \cite{georgiou} and evolutionary spectrum \cite{tj} methods.
\end{itemize}

\section{Conclusions}
\label{seven}
This paper has provided a comprehensive survey of the UAV channel characterization with measurement campaigns and statistical channel models. We have categorized the UAV channel measurement campaigns in low altitude platform based on the narrowband or wideband channel sounder, low\textendash cost and low\textendash power channel sounding solution, and widely deployed ground infrastructure. We have also reviewed empirical channel models for characterizing AG and AA propagation channel. Then we have classified the UAV channel modeling approaches as deterministic, stochastic and geometric\textendash stochastic models. Further, we have examined some challenging issues in the practicability of UAV communications related to airframe shadowing, the channel non\textendash stationarity, and diversity techniques. Finally we have presented some future research challenges which will be helpful to provide further insight of the UAV channel characterization for launching future measurement campaigns and proposing pragmatic framework for the effective UAV channel models.      


\end{document}